\documentclass[twocolumn,showpacs,preprintnumbers,amsmath,amssymb]{revtex4}
\usepackage{graphicx}% Include figure files
\usepackage{dcolumn}% Align table columns on decimal point
\usepackage{bm}% bold math

\begin{document}

\preprint{APS/123-QED}

\title{Contrasting Spin Dynamics Between Underdoped and Overdoped Ba(Fe$_{1-x}$Co$_{x}$)$_{2}$As$_{2}$}
% Force line breaks with \\

\author{F. L. Ning$^{1}$,  K. Ahilan$^{1}$, T. Imai$^{1,2}$, A. S. Sefat$^{3}$, M. A. McGuire$^{3}$, B. C. Sales$^{3}$, D. Mandrus$^{3}$, P. Cheng$^{4}$, B. Shen$^{4}$ and H.-H Wen$^{4}$}

\affiliation{$^{1}$Department of Physics and Astronomy, McMaster University, Hamilton, Ontario L8S4M1, Canada}
\affiliation{$^{2}$Canadian Institute for Advanced Research, Toronto, Ontario M5G1Z8, Canada}
\affiliation{$^{3}$Materials Science and Technology Division, Oak Ridge National Laboratory, TN 37831, USA}
\affiliation{$^{4}$National Laboratory for Superconductivity, Institute of Physics and Beijing National Laboratory for Condensed Matter Physics, Chinese Academy of Sciences, Beijing 100190, China}

\date{\today}% It is always \today, today,


\begin{abstract}
We report the first NMR investigation of spin dynamics  in the overdoped non-superconducting regime of Ba(Fe$_{1-x}$Co$_{x}$)$_{2}$As$_{2}$ up to $x =0.26$.  We demonstrate that the absence of inter-band transitions with large momentum transfer ${\bf Q}_{AF}\sim$~($\pi /a$, 0) between the hole and electron Fermi surfaces results in complete suppression of antiferromagnetic spin fluctuations for $x\gtrsim 0.15$.  Our experimental results provide direct evidence for a correlation between $T_c$ and the strength of ${\bf Q}_{AF}$ antiferromagnetic spin fluctuations.
\end{abstract}


\maketitle

%\section{\label{sec:level1}First-level heading:\protect\\ The line
%break was forced \lowercase{via} \textbackslash\textbackslash}

The critical temperature $T_c$ of the newly discovered iron-based superconductors \cite{Kamihara}  exceeds 50~K \cite{Norman,PhysicsToday}.  Intensive research efforts are under way world-wide to investigate the physical properties of these exciting new materials, yet the superconducting mechanism remains enigmatic.  The consensus reached so far is that the undoped parent phase of iron-arsenide superconductors (e.g. LaFeAsO and BaFe$_{2}$As$_{2}$) is a semi-metallic system with a SDW (Spin Density Wave) ordered ground state; upon doping a modest amount of electrons or holes, a high $T_c$ phase emerges from the magnetically ordered state \cite{Norman,PhysicsToday}.  Accordingly, it is natural to speculate that residual antiferromagnetic spin fluctuations (AFSF) associated with the SDW phase may be acting as the glue for the superconducting Cooper pairs.  

Unlike the case of the high $T_c$ cuprates, however, electrons in FeAs layers are always itinerant and there is no Mott insulating state in the electronic phase diagram.  Therefore a sensible theoretical approach is to choose uncorrelated itinerant electrons as the starting point, and crank up the electron-electron correlation effects.  On the other hand, on the experimental front, past studies exploring the possible relation between magnetism and superconductivity focused almost entirely on the underdoped side of the phase diagram near the SDW phase.  The evolution of the magnetic correlations on the overdoped side has been unexplored to date.

\begin{figure}[b]
\centering
\includegraphics[width=2.8in]{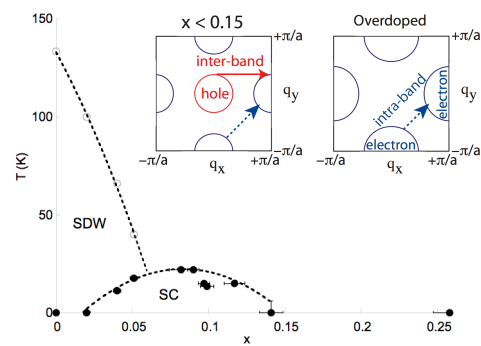}% Here is how to import EPS art
\caption{\label{Fig1:epsart} (Color online) The SDW and superconducting (SC) phase transition temperatures, $T_{SDW}$ and $T_c$, observed for our samples.  Also shown in the inset are the schematic representations of the Fermi surface geometry in the unfolded first Brillouin zone: the underdoped and optimally doped regimes $x < 0.15$ (left), and the overdoped non-superconducting regime (right).  Dashed and solid arrows represent the intra-band and inter-band transitions, respectively.  Filling of the hole Fermi surface by doped electrons results in the absence of ${\bf Q}_{AF}\sim$~($\pi /a$, 0) inter-band transitions between the hole and electron Fermi surfaces in the overdoped regime.   
}
\end{figure}

The primary goal of the present study is to fill this major void for the first time, and explore the magnetic correlation effects utilizing $^{75}$As NMR measurements in the overdoped region of the Ba(Fe$_{1-x}$Co$_{x}$)$_{2}$As$_{2}$ system \cite{Sefat,Ning1,Ning2,Ning3,Ahilan,Ni,Wen, Leyraud}.  Unlike the LaFeAsO$_{1-x}$F$_{x}$ system, one can transform Ba(Fe$_{1-x}$Co$_{x}$)$_{2}$As$_{2}$ into a non-superconducting metal by increasing the Co concentration above $x \sim 0.15$ \cite{Ni,Wen}.  Moreover, as a model system, Ba(Fe$_{1-x}$Co$_{x}$)$_{2}$As$_{2}$ has a major advantage over other iron-based superconductors  in a systematic investigation of electronic properties; one can conduct high precision measurements for homogeneous single crystals, and compare experimental results obtained by various techniques.  For example, recent ARPES \cite{Sekiba} and Hall  \cite{Wen} measurements in Ba(Fe$_{1-x}$Co$_{x}$)$_{2}$As$_{2}$ showed that overdoped electrons almost completely fill the hole Fermi surface at the center of the Brillouin zone when the doping level reaches $x\sim 0.15$, as schematically shown in Fig.1.  These findings imply that inter-band transitions with momentum transfer ${\bf Q}_{AF}\sim$~($\pi/a$, 0) between the hole and electron Fermi surfaces gradually disappear when the level of electron doping exceeds the optimal doping of $x \sim 0.08$ ($a$ is the distance between nearest neighbor iron sites).  How do the change of Fermi surface geometry and the absence of inter-band transitions affect spin fluctuations?  Is the absence of inter-band transitions the underlying cause of the suppression of superconductivity  in the overdoped region \cite{Sekiba,Wen}?  In what follows, we will demonstrate from our $^{75}$As NMR data that the filling of the hole Fermi surface results in complete suppression of AFSF.  Furthermore, we will show that the strength of spin fluctuations exhibits a clear correlation with $T_c$ in the overdoped regime above $x=0.08$.  Our findings suggest that AFSF associated with the inter-band transitions play a crucial role in the superconducting mechanism.

In Fig.2, we present representative field-swept $^{75}$As NMR lineshapes of the nuclear spin $I_{z}=+1/2$ to $-1/2$ central transition for single crystalline samples \cite{Wen} with $x=0.09$, 0.14, and 0.26.  The Co concentration $x$ and the superconducting critical temperature $T_c$ for each piece of crystal was determined from Energy Dispersive X-ray (EDX) measurements and in-plane resistivity $\rho_{ab}$, respectively, as summarized in Fig.1.  The sharp main peak in the NMR lineshape, As(0), arises from As sites with all four nearest neighbor (n.n.) sites occupied by Fe$^{2+}$ ions.  We also observe additional broad peaks for all concentrations, as reported earlier \cite{Ning1,Laplace,Julien}.  From systematic measurements of the NMR lineshapes at different magnetic fields, we found that the cause of the line splitting is second order nuclear quadrupole effects, and the Knight shifts of different peaks are comparable.   As shown in Fig.2, we can assign three additional peaks as As(1), As(2), and As(3) sites with 1, 2, and 3 of the n.n. Fe sites occupied by Co, because the intensity ratio is consistent with the probability of finding $N$~($=0-4$) Co at n.n. Fe sites, $P(N ; x) = C^{4}_{N} \cdot x^{N} \cdot (1-x)^{4-N}$.   We confirmed that spin dynamics measured at As(1) sites show qualitatively the same temperature and concentration dependencies as at As(0) sites.  We will discuss the complete details elsewhere, and focus our attention on As(0) sites in what follows.

\begin{figure}[b]
\centering
\includegraphics[width=3.2in]{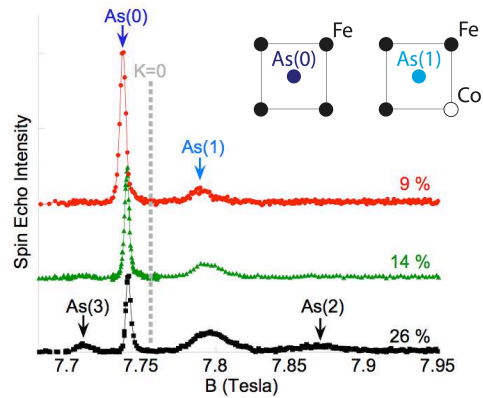}% Here is how to import EPS art
\caption{\label{Fig2:epsart} (Color online) Field swept $^{75}$As NMR lineshapes of overdoped Ba(Fe$_{1-x}$Co$_{x}$)$_{2}$As$_{2}$ with $x=0.09$ at 20~K ($>T_{c}$), $x=0.14$ at 4.2~K ($>T_{c}$), and non-superconducting $x=0.26$ at 4.2~K with external magnetic field $B$ // c-axis.  The NMR frequency is fixed at $f = 56.555$~MHz.  The grey dashed line marks the expected resonance position for $K=0$.  Notice the systematic increase in the relative intensity of the As(N) peaks ($N=1 - 3$) for larger $x$.  Inset : schematic representations of the Fe and Co coordinations of As(0) and As(1) sites.}
\end{figure}

In Fig.3, we present the temperature dependence of the NMR Knight shift $K$ in overdoped $x=0.09$, 0.12, 0.14 and 0.26 samples measured for the main As(0) sites.  For comparison, we also present our earlier results for optimum and underdoped samples $x\leq 0.08$ \cite{Ning2,Ning1}.  In the metallic state above $T_{SDW}$ and $T_{c}$, all compositions exhibit qualitatively the same behavior; $K$ decreases monotonically with decreasing temperature, then levels off below $\sim 50$~K.   NMR Knight shift is related to the local electron spin susceptibility $\chi_{spin}$ by $K = K_{spin} + K_{chem}$; $K_{spin} = A_{hf}\chi_{spin}/N_{A}\mu_{B}$ is the spin contribution to the Knight shift, where $A_{hf} = 18.8$~kOe/$\mu_{B}$ \cite{Kitagawa} is the hyperfine coupling constant between $^{75}$As nuclear spins and surrounding electrons, $N_{A}$ is Avogadro's number, and $\mu_{B}$ is the Bohr magneton.  The temperature independent chemical shift is $K_{chem} \sim 0.22$\% for $x=0.08$ \cite{Ning1}, but has a small concentration dependence, as shown below.  Our results in Fig.3 indicate that $\chi_{spin}$ shows qualitatively the same behavior for all compositions regardless of the nature of the ground state.

\begin{figure}
\centering
\includegraphics[width=3.2in]{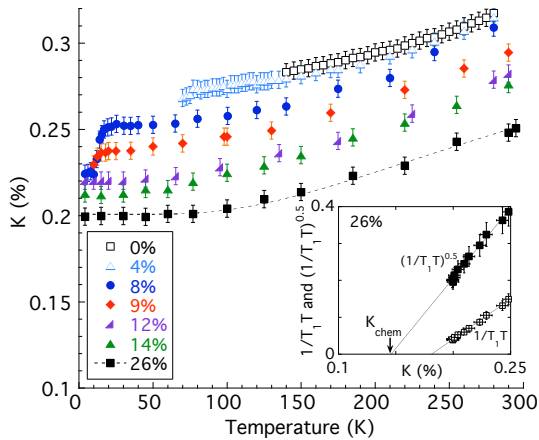} \\% Here is how to import EPS art
\caption{\label{Fig3:epsart} (Color Online) ${\bf q}={\bf 0}$ uniform susceptibility as measured by $^{75}$As NMR Knight shift $K$ for As(0) sites in representative compositions.  The dashed curve is a fit of $x=0.26$ data to an activation form, $K = 0.20 + 0.23 \times exp (-\Delta/T)$, with $\Delta/k_{B} = 450$~K.  Inset :  $(1/T_{1}T)^{0.5}$ and $1/T_{1}T$ for $x=0.26$ plotted as a function of  $K$ with temperature as the implicit parameter.
}
\end{figure}

\begin{figure}
\centering
\includegraphics[width=3.2in]{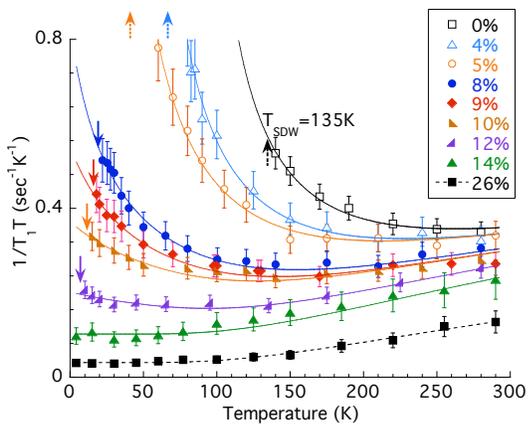}% Here is how to import EPS art
\caption{\label{Fig4:epsart} (Color Online) $1/T_{1}T$ measured at As(0) sites for various concentrations $x$ with magnetic field $B$ applied along the ab-plane.  Solid and dashed arrows mark $T_{c}$ and $T_{SDW}$, respectively.  Solid and dashed curves are the best fits with (for $x \leq 0.14$) and without (for $x=0.26$) a Curie-Weiss term arising from AFSF; see the main text for details.
}
\end{figure}

However, the qualitative similarity observed for $K$ must not be mistaken as evidence for overall similarity of spin excitations between $x=0$ and $x=0.26$.  After all, $K$ probes only the uniform ${\bf q}={\bf 0}$ wave vector mode of the spin susceptibility, $\chi_{spin}$.  In order to see the influence of doping on spin excitations, it is more useful to look into the nuclear spin-lattice relaxation rate $1/T_{1}$ divided by $T$ (i.e. $1/T_{1}T$) presented in Fig.4.  $1/T_{1}T$ measures the ${\bf q}$ integral of the imaginary part of the dynamical spin susceptibility, $\chi"({\bf q}, f)$, in the first Brillouin zone, i.e. $1/T_{1}T \propto \sum_{{\bf q}} |A_{hf}({\bf q})|^{2} \chi"({\bf q}, f)/f$, where $f \sim 56.5$~MHz is the NMR frequency.  It is important to note that $1/T_{1}T$ reflects the summation of all different ${\bf q}$~modes of  spin fluctuations, i.e. both inter-band spin excitations with large momentum transfer $\Delta {\bf q} \sim {\bf Q}_{AF}$ {\it and} intra-band spin excitations with smaller momentum transfers.  

We start our discussion on the evolution of spin excitations from the non-superconducting metallic phase at $x=0.26$.  A crucial difference between $x=0.26$ and the optimally doped superconductor $x=0.08$ is that $1/T_{1}T$ of the former levels off to a very small constant value below $\sim 50$~K.  We recall that, within a canonical Fermi liquid picture, $1/T_{1}T \propto N(E_{F})^{2}$ due to Fermi's golden rule (where $N(E_{F})$ is the density of states at the Fermi energy).  On the other hand, $K_{spin} \propto \chi_{spin}  = \mu_{B}^{2} N(E_{F})$ from Pauli spin susceptibility.  Accordingly, the Korringa relation, $(1/T_{1}T)^{0.5} = (constant) \times K_{spin}$, is a benchmark test for the applicability of the Fermi liquid theory to a strongly correlated electron system.  Plotted in the inset to Fig.3 is $(1/T_{1}T)^{0.5}$ as a function of $K$, where temperature has been chosen as the implicit parameter.  We find a good linear relation between these two quantities for the whole temperature range between 4.2~K and 290~K.  This means that, when only intra-band electron excitations exist, the nature of spin excitations in $x=0.26$ is consistent with a Fermi liquid picture.  Our finding is also consistent with the fact that in-plane resistivity varies as $\rho_{ab} \sim T^{2}$ in $x=0.26$ \cite{Ni,Wen}, another benchmark for Fermi liquid behavior.  We estimate $K_{chem}=0.15$~\% from the extrapolation of the linear fit to the $(1/T_{1}T)^{0.5}$ vs. $K$ plot.  The net spin contribution to the Knight shift below 50~K can then be determined as $K_{spin} = K - K_{chem} =0.2-0.15= 0.05$~\%, hence $\chi_{spin} \sim 1.5 \times 10^{-4}$~emu/mol-Fe.  ($K_{chem}$ may be slightly underestimated because we ignored possible small orbital contributions to $1/T_{1}T$, hence $\chi_{spin}$ may be slightly overestimated.)  According to LDA band calculations, the bare density of states $N_{o}(E_{F}) \sim 4.6$~eV$^{-1}$/f.u. in Ba(Fe$_{1-x}$Co$_{x}$)$_{2}$As$_{2}$ \cite{Sefat,Singh}, hence we expect bare Pauli spin susceptibility $\chi_{spin}^{band} \sim 0.8 \times 10^{-4}$~emu/mol-Fe. The factor of $\sim 2$ enhancement of $\chi_{spin}$ over $\chi_{spin}^{band}$ may be the consequence of mild mass enhancement of electrons due to electron-electron interactions.  We caution, however, that we also found a linear relation between $1/T_{1}T$ and $K$ as presented in the inset to Fig.3.  In fact, we can fit both $1/T_{1}T$ and $K$ of the $x=0.26$ sample with the same empirical activation form, $\alpha + \beta \cdot exp(-\Delta / k_{B}T)$, and a common phenomenological gap $\Delta/k_{B} \sim 450\pm 40$~K, as shown by the dashed curves in Fig.3 and Fig.4.  This might be an indication that spin excitations in the overdoped metallic phase are still dominated by over-damped paramagnons.  In this scenario, we obtain $K_{chem}\sim 0.18$~\% from the inset to Fig.3, and $\chi_{spin} \sim 0.6 \times 10^{-4}$~emu/mol-Fe.      

How do spin excitations evolve when we reduce the level of electron doping below $x \sim 0.26$?  We recall that a hole pocket will begin to grow once we reduce the doping level below $x \sim 0.15$ \cite{Sekiba,Wen}.  This means that if the presence of the hole Fermi surface is playing a crucial role in the spin excitations in the superconducting regime below $x\sim 0.15$, we may find a qualitative change in spin excitations below this concentration.  In fact, our results in Fig.4 show that $1/T_{1}T$ exhibits an upturn for $x\leq0.12$ due to the growth of AFSF.  Further reduction of the doping level results in divergence of $1/T_{1}T$ toward $T_{SDW}$ due to that of AFSF with $\Delta {\bf q} \sim {\bf Q}_{AF}$ ($1/T_{1}T$ does not blow up at $T_{SDW}=135$~K for the undoped $x=0$ sample, because the SDW transition is first order for $x=0$ \cite{Kitagawa}).  

We can see the systematics more clearly by plotting the concentration $x$ dependence of $1/T_{1}T$ observed at 25~K ($\gtrsim T_c$ of $x=0.08$), as shown in Fig.5a.   The strength of spin fluctuations at 25~K, as reflected by the magnitude of $1/T_{1}T$, shows only a mild concentration dependence from $x=0.26$ down to $x \sim 0.15$, but grows dramatically below $x \sim 0.15$.  Equally interesting is the fact that the growth of spin fluctuations with decreasing $x$ correlates with that of $T_c$ in Fig.1.    Thus our $1/T_{1}T$ data clearly establish that (a) robust AFSF remain even in the optimum ($x=0.08$) and slightly overdoped ($x=0.09-0.10$) superconducting samples, and (b) it is unlikely that AFSF and the superconducting mechanism compete with each other.  If the presence of AFSF was genuinely detrimental to the formation of superconducting Cooper pairs, the $x=0.08$ sample with strong enhancement of AFSF below $\sim 100$~K would not have the maximum $T_c$.

\begin{figure}
\centering
\includegraphics[width=3.4in]{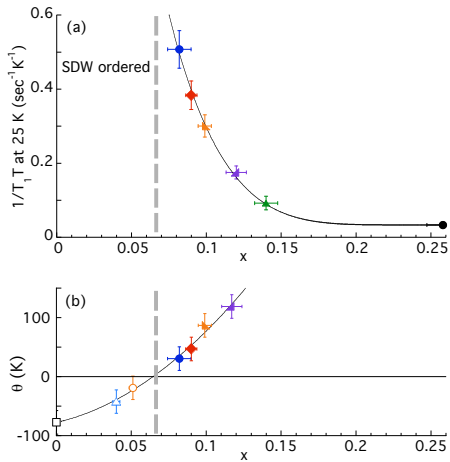}% Here is how to import EPS art
\caption{\label{Fig4:epsart} (Color online) The concentration dependence of (a) the strength of paramagnetic spin fluctuations as measured by $1/T_{1}T$ observed at 25~K, and (b) Weiss temperature $\theta$ obtained from the fit in Fig.4.  Solid curves are guides for the eyes.}
\end{figure}

In order to gain additional insight into the relation between AFSF and superconductivity, we fit the $1/T_{1}T$ data with a simple phenomenological two-component model, $1/T_{1}T = (1/T_{1}T)_{inter} +  (1/T_{1}T)_{intra}$, where we represent the contributions of the inter-band AFSF with a Curie-Weiss term, $(1/T_{1}T)_{inter} = C/(T + \theta)$.  Since the temperature dependence of $1/T_{1}T$ above $\sim 150$~K is similar for a broad concentration range, it is reasonable to assume that the intra-band contributions may be represented by the same phenomenologocal activation form, $(1/T_{1}T)_{intra}=\alpha + \beta \cdot exp(-\Delta /k_{B}T)$, employed earlier for $x=0.26$.  We take the same $\Delta /k_{B}=450$~K for all compositions as determined from the fit of $x=0.26$, since the Knight shift data show nearly identical temperature dependence except for constant offsets.  For simplicity, we also fix the constant $\alpha$~($=0.11$) and $\beta$~($=0.63$) from the best fit of the $1/T_{1}T$ data for $x=0.14$ sample.  In principle, $(1/T_{1}T)_{intra}$ may be slightly concentration dependent below $x=0.14$; however, we found that floating the values of $\alpha$, $\beta$ and $\Delta$ does not alter the essential conclusions, because $(1/T_{1}T)_{inter}$ is the dominant contribution for $x \leq 0.1$.

Despite the simplicity of our minimalist model, the fits presented in Fig.4 capture the essential aspects of the temperature and concentration dependences of our $1/T_{1}T$ data remarkably well for all compositions.  The resulting value of the Weiss temperature $\theta$ is summarized in Fig.5b.  The negative value of $\theta$ for $x\leq 0.05$ implies that these samples are gradually approaching a magnetic instability from $T >> T_{SDW}$.  On the other hand, the relatively large positive value of  $\theta \sim 119$~K for $x=0.12$ reflects the fact that the overdoped sample is far from magnetic instabilities, hence the growth of AFSF is only modest.  The small positive value of $\theta = 31$~K for $x=0.08$ is evidence for the close proximity of the optimally doped superconducting phase with a magnetic instability, i.e. high $T_c$ superconductivity is realized near a quantum critical point, where we expect $\theta =0$.  In passing, $C = 24 \pm 4$~sec$^{-1}$ is independent of $x$ from $x=0$ to $x=0.1$, then decreases to $C \sim 12$~sec$^{-1}$ for $x=0.12$ and $C \sim 0$~sec$^{-1}$ for $x=0.14$.  That is, the contribution of  the Curie-Weiss term associated with inter-band transitions  becomes negligibly small for $x=0.14$.  We also note that we arrive at analogous conclusions even if we employ the $1/T_{1}T$ data measured with magnetic field $B$ applied along the c-axis \cite{Ning2}.

To summarize, we have investigated the spin excitations of Ba(Fe$_{1-x}$Co$_{x}$)$_{2}$As$_{2}$ over the entire doping range for the first time.  Our NMR data for the overdoped metallic phase $x=0.26$ is consistent with the Korringa relation expected for canonical Fermi liquid systems.  However, as we decrease the level of doping across $x \sim 0.15$, where a hole Fermi surface emerges in the center of the Brillouin zone, we find a dramatic enhancement of ${\bf Q}_{AF}\sim$~($\pi/a$, 0) antiferromagnetic spin fluctuations associated with inter-band transitions.  The superconducting critical temperature $T_c$ is optimized when these spin fluctuations are modestly enhanced, to the extent that SDW ordering does not set in.  The correlation observed between the strength of  ${\bf Q}_{AF}$ antiferromagnetic spin fluctuations and $T_c$ suggests the former plays a crucial role in the superconducting mechanism.    

The work at McMaster was supported by NSERC, CFI, and CIFAR.  Research at ORNL was sponsored by the Division of Materials Sciences and Engineering, Office of Basic Energy Sciences, U.S. Department of Energy.  The work at Beijing was supported by NSF, the Ministry of Science and Technology of China, and the Chinese Academy of Sciences.\\

%Just because of unusual number of tables stacked at end

%\bibliography{Ning_v2}

\end{document}